\documentclass[english,12pt,a4paper]{article}
\usepackage{a4wide}

\usepackage{graphicx}
\graphicspath{
                          {./figures/}
                          {./}
                         }
\DeclareGraphicsExtensions{.pdf,.png}

\usepackage{subfig}
\usepackage{lscape}
\usepackage{babel}

\usepackage{abstract}

\usepackage{multirow}
\usepackage{amsmath}
\usepackage{amssymb}
\usepackage[usenames,dvipsnames,pdftex]{color}

\usepackage{authblk}

\usepackage{bm}

\usepackage[numbers,sort&compress]{natbib}

\usepackage{siunitx}
\usepackage{booktabs}
\usepackage{miller}

\usepackage{cleveref}

\begin{document}

\title{Computing forces on interface elements exerted by dislocations in an elastically anisotropic crystalline material}

\author[1,*]{B.~Liu}
\author[1]{A.~Arsenlis}
\author[1]{S.~Aubry}
\affil[1]{\normalsize{Lawrence Livermore National Laboratory, Livermore, CA 94550, USA}}
\affil[*]{\normalsize{Corresponding author: bingliu\textcircled{a}llnl.gov (B.~Liu)}}

\maketitle

\begin{abstract}
Driven by the growing interest in numerical simulations of dislocation--interface interactions in general crystalline materials with elastic anisotropy, we develop algorithms for the integration of interface tractions needed to couple dislocation dynamics with a finite element or boundary element solver. The dislocation stress fields in elastically anisotropic media are made analytically accessible through the spherical harmonics expansion of the derivative of Green's function, and analytical expressions for the forces on interface elements are derived by analytically integrating the spherical harmonics series recursively. Compared with numerical integration by Gaussian quadrature, the newly developed analytical algorithm for interface traction integration is highly beneficial in terms of both computation precision and speed.
\end{abstract}

\section{Introduction}

It is well known that a dislocation approaching a free surface experiences an attractive force, so called image force \cite{Hull&Bacon2011}. In multiphase materials or single-phase elastically anisotropic polycrystalline materials, a dislocation near a phase or grain boundary is subject to a similar force due to the change in elasticity across the interface, which can be either attractive or repulsive \cite{Hirth1972, Sutton&Balliffi1995, Priester2013}.  As the magnitudes of these forces are inversely proportional to the dislocation--interface distance, the effect of free surfaces on dislocation motion and multiplication becomes significant in submicrometre-sized crystals \cite{Shan2008, Oh2009, Brinckmann2008, Weinberger&Cai2008}, and the elastic interactions between dislocations and phase/grain boundaries have a stronger impact on the mechanical properties of the nanostructured materials \cite{Gleiter2000, Jin2006, Estrin2011, Jang2012} than those of their coarse-grained counterparts.

For free surfaces and phase/grain boundaries of elastically anisotropic half spaces, such virtual forces (negative energy gradients) can be determined through the image force theorem of Barnett and Lothe~\cite{Barnett&Lothe1974, Belov1983,Khalfallah1993, Priester&Khalfallah1994, Khalfallah&Priester1999}. For finite domains of more complex shapes, this problem of elastic interactions between dislocations and interfaces can only be solved through the coupling of a dislocation dynamics (DD) code and a finite element (FE) or boundary element (BE) solver \cite{vanderGiessen&Needleman1995, Zbib2002, Weygand2002, ODay&Curtin2004, ODay&Curtin2005, Tang2006, El-Awady2008, Deng2008, Shishvan2011, Vattre2014, Crone2014}. In such cases, the stress field of the dislocation in an infinite elastically homogeneous medium is used to calculate the traction on the surface bounding the elastic solid, and then correction fields are added to impose the proper boundary conditions on the domain.  For free surfaces this results in an imposition of an equal and opposite surface traction such that the net result is a zero traction condition, and for phase/grain boundaries there is a traction balance and displacement continuity condition that must be imposed. This work is focused on the first part of DD-FE/BE simulations of dislocation--interface interactions, i.e. determination of forces on interface elements due to tractions imposed by dislocations in an infinite elastically anisotropic medium.

Interface tractions exerted by dislocation stress fields must be integrated into nodal forces in a FE or BE solver. These nodal forces are surface integrals of the traction field ${\bm T}$ (force per unit area) over the individual interface FE or BE elements \cite{Weygand2002, El-Awady2008, Crone2014}, 
\begin{equation}
{\bm F}^{(n)} = \int_{S}{\bm T}\left({\bm x}\right)N_{S}^{(n)}({\bm x})\mathrm{d}S = \int_{S}\left[{\bm \sigma}\left({\bm x}\right)\cdot{\bm n}\right]N_{S}^{(n)}({\bm x})\mathrm{d}S,
\end{equation}
where ${\bm F}^{(n)}$ is the force on a FE or BE node $n$ of an interface element $S$ exerted by dislocation stress field ${\bm \sigma}\left({\bm x}\right)$, ${\bm n}$ is the interface normal, and $N_{s}^{(n)}({\bm x})$ are the FE or BE shape functions. In this paper, we will refer to these nodal forces as traction forces.  The traction force on a FE or BE node is analogous to the interaction force due to dislocation stress field on a dislocation node in nodal based (one-dimensional FE) dislocation dynamics models \cite{ Schwarz1999, Ghoniem2000, Weygand2002, Arsenlis2007}. The dislocation interaction forces are line integrals of the Peach-Koehler force ${\bm f}^{PK}$ (force per unit length) along the individual dislocation segments \cite{Weygand2002, Arsenlis2007},
\begin{equation}
{\bm f}^{(n)} = \int_{L}{\bm f}^{PK}\left({\bm x}\right)N_{L}^{(n)}({\bm x})\mathrm{d}L = \int_{L}\left[{\bm \sigma}\left({\bm x}\right)\cdot{\bm b}\times{\bm t}\right]N_{L}^{(n)}({\bm x})\mathrm{d}L,
\end{equation}
where ${\bm f}^{(n)}$ is the force on a node $n$ of a dislocation segment $L$ exerted by dislocation stress field ${\bm \sigma}\left({\bm x}\right)$, with ${\bm b}$, ${\bm t}$, and $N_{L}^{(n)}({\bm x})$ being the Burgers vector, line direction, and shape function of the dislocation segment, respectively. While the dislocation interaction force calculation is the core of a DD model, the interface traction force evaluation is the main link in DD-FE/BE finite domain simulations. However, the compromise between accuracy and efficiency of the numerical integrations of dislocation interaction forces and surface traction forces has been the bottleneck for large-scale DD simulations \cite{Arsenlis2007} and DD--FE/BE simulations of the elastic interactions between dislocations and free surfaces \cite{Liu2000, Weygand2002, Weinberger2007, El-Awady2008, Fertig&Baker2009}. Due to the limitations of numerical integrations, an alternative analytical integration of interface traction forces is highly desirable.
  

For isotropic elastic media, the stress field of a dislocation loop can be calculated analytically through line integrations of the derivatives of Green's function along piecewise straight dislocation segments \cite{Hirth&Lothe1982, Cai2006}. Arsenlis et al.~\cite{Arsenlis2007} developed analytical expressions for dislocation interaction force calculations that involves double integrations along the individual pairs of interacting dislocation segments, i.e. the first line integration along the source segment to get its stress field and the second line integration along the receiving segment to obtain the interaction force.  These analytical expressions for dislocation interaction force calculations have already been used in many DD simulations, e.g. discovery of ternary dislocation junctions in body-centered cubic metals \cite{Bulatov2006}, interpretation of the size-dependent strength for micrometer-sized crystals \cite{Rao2008}, determination of low-angle grain boundary penetration resistances \cite{Liu2011,Liu2012}, observation of strain localization via defect-free channels in highly irradiated materials \cite{Arsenlis2012}, and revealing the mechanisms of dynamic recovery during high temperature creep of single-crystal superalloys \cite{Liu2014}. Queyreau et al.~\cite{Queyreau2014} have recently formulated analytical expressions to calculate surface traction forces induced by stress field of a dislocation in isotropic elastic media for rectangular surface elements, which give the precise solutions to triple integrals, i.e. one integral along the dislocation segments for the stress field and then a double integral over the surface element to obtain the traction force.

In anisotropic elasticity theory of dislocations, the analytical expression for the stress field of an arbitrary dislocation loop does not exist \cite{Bacon1980, Chu2012}. The dislocation stress field calculation based on Stroh's formulism \cite{Stroh1962, Hirth&Lothe1982} requires numerical integrations \cite{Rhee2001, Yin2010} or solving an eigenvalue problem for a six by six matrix \cite{Yin2010} to obtain the associated matrices and their angular derivatives. The dislocation stress field computations using Mura's formula \cite{Mura1987} has relied on direct numerical integrations of the derivatives of Green's function \cite{Han2003}. Recently, Aubry and Arsenlis \cite{Aubry&Arsenlis2013} used a truncated spherical harmonics expansion to approximate the derivatives of Green's function, and have analytically integrated the spherical harmonics series to calculate dislocation stress field (single integrals) and interaction force (double integrals) for straight dislocation segments. In this work, we use the same spherical harmonics expansion of the derivatives of Green's function formulated in the previous work \cite{Aubry&Arsenlis2013}, and analytically integrate the associated spherical harmonics series to determine the interface traction force (triple integrals) in anisotropic elastic media for quadrilateral surface elements.


\section{Method} 

In a homogeneous infinite linear elastic solid, the stress field of a dislocation loop can be expressed in terms of a contour integral along the loop \cite{Mura1987}, 
\begin{equation}
\sigma_{js}\left({\bm x}\right) = \epsilon_{ngr}C_{jsvg}C_{pdwn}b_{w}\oint_{L}\frac{\partial G_{vp}}{\partial x_{d}}\left({\bm x}-{\bm x}'\right)\mathrm{d}x'_r,
\end{equation}
where $C_{ijkl}$ is the elastic stiffness tensor, $\epsilon$ is the permutation tensor, ${\bm b}$ is the Burgers vector of the dislocation loop, and ${\partial G_{vp}}/{\partial x_{d}}$ is the derivative of the Green's function $G_{vp}\left({\bm x}-{\bm x}'\right)$, which is defined as the displacement in the $x_v$-direction at point ${\bm x}$ in response to a unit point force in the $x_p$-direction applied at point ${\bm x}'$. 

\begin{figure}
\centering
\includegraphics[width=0.5\linewidth]{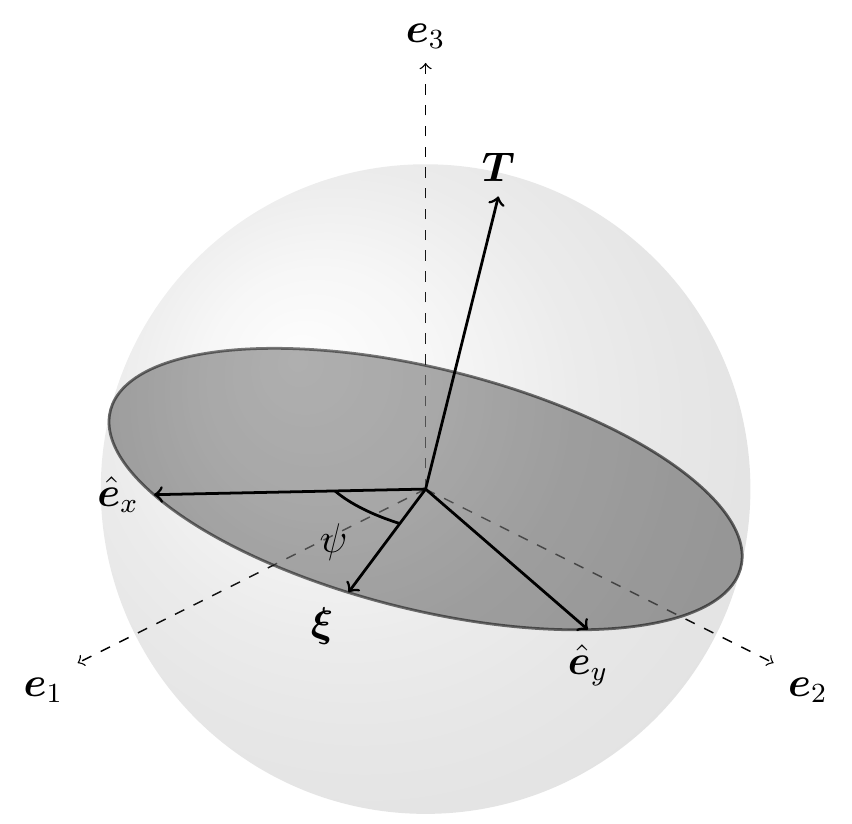}
\caption{Unit sphere in anisotropic elasticity}
\label{fig:GreenFunctionGeo}
\end{figure}

The Green's function in an anisotropic elastic medium has been obtained as a single integral \cite{Bacon1980},
\begin{equation}
G_{vp}=\frac{1}{4 \pi^{2} R}\int^{\pi}_{0}M^{-1}_{vp}\left(\xi\right)\mathrm{d}\psi,
\end{equation}
where
\begin{equation*}
M^{-1}_{vp}\left(\xi\right)=\frac{\epsilon_{vsm}\epsilon_{prw}\left(\xi\xi\right)_{sr}\left(\xi\xi\right)_{mw}}{2\epsilon_{lgn}\left(\xi\xi\right)_{1l}\left(\xi\xi\right)_{2g}\left(\xi\xi\right)_{3n}},
\end{equation*}
with the notation $\left(\xi\xi\right)_{ij}=\xi_{k}C_{kijl}\xi_{l}$. $R$ is the norm of vector ${\bm R}={\bm x}-{\bm x}'$, i.e. $R=\|{\bm R}\|$. ${\bm T}$ is the direction of vector ${\bm R}$, i.e. ${\bm T}={\bm R}/R$. $\xi$ is a unit vector that varies in the plane ${\bm \xi} \cdot {\bm T}=0$, $\hat{\bm e}_{x}$ and $\hat{\bm e}_{y}$ are two orthogonal unit vectors in the same plane, and the angle between $\hat{\bm e}_{x}$ and $\xi$ is $\psi$, Fig.~\ref{fig:GreenFunctionGeo}.

The corresponding integral expression for the derivative of Green's function is \cite{Barnett1972},
\begin{equation}
\frac{\partial G_{vp}}{\partial x_{d}}= \frac{1}{4 \pi^{2} R^{2}}\int^{\pi}_{0} \left(-T_{d}M^{-1}_{vp}+\xi_{d}N_{vp}\right)\mathrm{d}\psi,
\end{equation}
where $N_{vp}=C_{jrnw}M^{-1}_{vj}M^{-1}_{np}\left(\xi_{r}T_{w}+\xi_{w}T_{r}\right)$. See the overview of Bacon et al.~\cite{Bacon1980} for more details. 

The derivative of the Green's function is a product of a part depending only on $1/R^2$ and an angular part ${\bm g}$ depending only on the direction ${\bm T}$
\begin{equation}
g_{vpd}\left({\bm T}\right) = g_{vpd}\left(\theta, \phi\right) = \int^{\pi}_{0} \left(-T_{d}M^{-1}_{vp}+\xi_{d}N_{vp}\right)\mathrm{d}\psi,
\end{equation}
where $\left(\theta, \phi\right)$ are the spherical coordinates of ${\bm T}$.

There is no analytical expression for $g_{vpd}$, but the function $g_{vpd}\left({\bm T}\right)$ is suitable for decomposition in spherical harmonics.
\begin{equation}
{\bm g}\left({\bm T}\right)=\sum^{\infty}_{l=0}\sum^{l}_{m=-l}{\bm g}^{lm}Y^{m}_{l}\left({\bm T}\right)
\end{equation}
The expansion coefficients ${\bm g}^{lm}$ are independent of ${\bm T}\left(\theta,\phi\right)$, and are defined as
\begin{equation}
{\bm g}^{lm}=\int^{2\pi}_{0}\int^{\pi}_{0}{\bm g}\left(\theta,\phi\right)Y^{m\ast}_{l}\left(\theta,\phi\right)\sin\theta\mathrm{d}\theta\mathrm{d}\phi
\end{equation}
The spherical harmonics $Y^{m}_{l}$ are defined as the complex functions
\begin{equation}
Y^{m}_{l}(\theta,\phi)=\sqrt{\frac{2l+1}{4\pi}\frac{(l-m)!}{(l+m)!}} \, P^{m}_{l}(\cos\theta) \, e^{im\phi},
\end{equation}
where $P^{m}_{l}$ are the associated Legendre polynomials.
To be consistent with the definition of elastic stiffness tensor $C_{ijkl}$, we rewrite $Y^{m}_{l}$ in the Cartesian coordinate system $\left({\bm e}_1, {\bm e}_2, {\bm e}_3\right)$, i.e. $x={\bm T}\cdot{\bm e}_1$,  $y={\bm T}\cdot{\bm e}_2$, and $z={\bm T}\cdot{\bm e}_3$, in the form of
\begin{equation}
Y^{m}_{l}(x,y,z)= f_{m}(x,y) \sum^{\left[\left(l- |m|\right)/2\right]}_{k=0}\bar{Q}^{|m|}_{l}(k) z^{l-|m|-2k}
\end{equation}
where
\begin{equation*}
f_{m}(x,y) = \left\{
\begin{array}{l l}
(x+iy)^m & m \geq 0 \\
(x-iy)^{-m} & m < 0
\end{array}\right.
\end{equation*}
and
\begin{equation*}
\bar{Q}^{m}_{l}(k)=\frac{(-1)^{m+k}}{4 \pi^{2}}\frac{m!}{2^{l}}\sqrt{\frac{2l+1}{4\pi}\frac{(l-m)!}{(l+m)!}}
\left(\begin{array}{c} l \\ k \end{array}\right)
\left(\begin{array}{c} 2l-2k \\ l \end{array}\right)
\left(\begin{array}{c} l-2k \\ m \end{array}\right)
\end{equation*}

The function ${\bm g}$ can then be defined as
\begin{equation}
{\bm g}(x,y,z)=\sum^{\infty}_{l=0}\sum^{l}_{m=0} \Re\left((x+iy)^{m}{\bm g}^{lm}\right) \sum^{[(l-m)/2]}_{k=0} Q^{m}_{l}(k) z^{l-m-2k}
\end{equation}
where we note $\Re(x)$ is the real part of $x$, $Q^{0}_{l}(k)=\bar{Q}^{0}_{l}(k)$ when $m=0$, and $Q^{m}_{l}(k)=2\bar{Q}^{m}_{l}(k)$ when $m>0$.

Using the spherical harmonics series expansion of its angular part and defining ${\bm e}_{12}={\bm e}_1 + i{\bm e}_2$ with $x+iy=\frac{\bm R}{R}\cdot{\bm e}_{12}$ and $z=\frac{\bm R}{R}\cdot{\bm e}_{3}$, the derivative of the Green's function can be evaluated by
\begin{equation}
\frac{\partial G_{vp}}{\partial x_{d}}({\bm R})=\sum^{\infty}_{l=0}\sum^{l}_{m=0}\sum^{[(l-m)/2]}_{k=0}\Re\left(Q^{m}_{l}(k)g^{lm}_{vpd}\frac{({\bm R}\cdot{\bm e}_{12})^{m}({\bm R}\cdot{\bm e}_{3})^{l-m-2k}}{R^{l-2k+2}}\right).
\end{equation}
This definition involves a quotient of terms that are a function of $R$, which depends only on
two variables $m$ and $l-2k$. It can be simplified further to obtain
\begin{equation}
\frac{\partial G_{vp}}{\partial x_{d}}({\bm R})=\sum^{\infty}_{q=0}\sum^{2q+1}_{m=0}\ \Re\left(S^{qm}_{vpd}\frac{({\bm R}\cdot{\bm e}_{12})^{m}({\bm R}\cdot{\bm e}_{3})^{2q+1-m}}{R^{2q+3}}\right)
\end{equation}
where $S^{qm}_{vpd}$ is a sum of products composed of $Q^{m}_{l}(k)$ and $g^{lm}_{vpd}$. The Green's function and its derivatives depend only on odd powers of $1/R$. This property means that in the spherical harmonics expansion, the non-zero terms correspond to the odd powers of $1/R$.

Within linear elasticity, the total stress field of a dislocation network is a superposition of the stress fields of individual dislocation segments in the network. For a straight dislocation segment linking two nodes at Cartesian coordinates ${\bm x}_1$ and ${\bm x}_2$, respectively, its stress field in an anisotropic elastic medium can then be expressed by
\begin{eqnarray}
\sigma_{js}\left({\bm x}\right) = \epsilon_{ngr}C_{jsvg}C_{pdwn}b_{w}\sum^{\infty}_{q=0}\sum^{2q+1}_{m=0}\Re\left(S^{qm}_{vpd}\int^{{\bm x}_2}_{{\bm x}_1}\frac{\left({\bm R}\cdot{\bm e}_{12}\right)^{m}\left({\bm R}\cdot{\bm e}_3\right)^{2q+1-m}}{R^{2q+3}}\mathrm{d}x'_r\right)
\label{eq:stressField}
\end{eqnarray}

\begin{figure}
\centering
\includegraphics[width=0.66\linewidth]{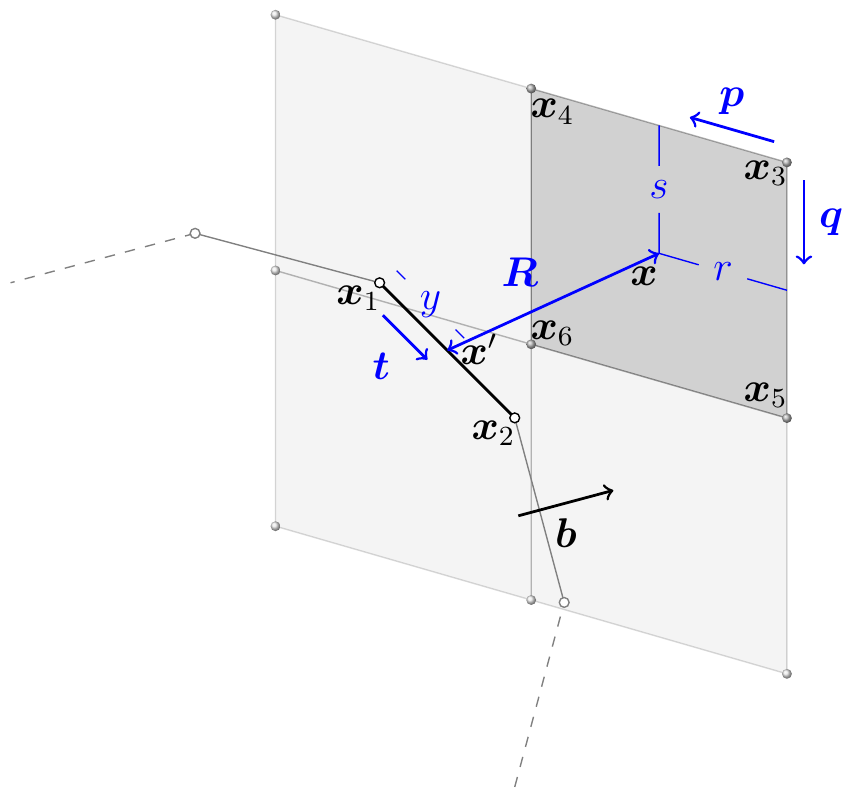}
\caption{Geometry and associated variables for interface traction integration}
\label{fig:DislocationTractionGeo}
\end{figure}

Consider a quadrilateral element delimited by its Cartesian coordinates ${\bm x}_3$, ${\bm x}_4$, ${\bm x}_5$,  and ${\bm x}_6$ as shown in Fig.~\ref{fig:DislocationTractionGeo},  where ${\bm x}$ is the coordinate that spans within the element, and ${\bm x}'$ is the coordinate that spans along the dislocation segment delimited by ${\bm x}_1$ and ${\bm x}_2$. The vector ${\bm R}={\bm x}-{\bm x}'$ can be rewritten as ${\bm R}=y{\bm t}+r{\bm p}+s{\bm q}$, with ${\bm t}=\frac{{\bm x}_2-{\bm x}_1}{\| {\bm x}_2-{\bm x}_1\|}$, ${\bm p}=\frac{{\bm x}_4-{\bm x}_3}{\| {\bm x}_4-{\bm x}_3\|}$, and ${\bm q}=\frac{{\bm x}_5-{\bm x}_3}{\| {\bm x}_5-{\bm x}_3\|}$.

Using ${\bm n}=\frac{{\bm p}\times{\bm q}}{\|{\bm p}\times{\bm q}\|}$ and $\mathrm{d}S=\|{\bm p}\times{\bm q}\|\mathrm{d}r\mathrm{d}s$, with a linear shape function in the form of a four-term polynomial $N^{(n)}({\bm x})=a_{0}+a_{1}r+a_{2}s+a_{3}rs$, the traction force ${\bm F}^{(n)}$ can be expressed as
\begin{eqnarray}
{\bm F}^{(n)} = \int^{s_2}_{s_1}\int^{r_2}_{r_1}{\bm \sigma}({\bm x})\cdot\left({\bm p}\times{\bm q}\right)(a_{0}+a_{1}r+a_{2}s+a_{3}rs)\mathrm{d}r\mathrm{d}s
\label{eq:tractionForce}
\end{eqnarray}

Combining Eq.~\eqref{eq:stressField} and Eq.~\eqref{eq:tractionForce} with $\mathrm{d}{\bm x}' = -{\bm t}\mathrm{d}y$, the interface nodal force due to dislocation traction in an anisotropic elastic medium can be determined through
\begin{eqnarray}
\begin{array}{l l l}
{F}^{\left(n\right)}_j & = & - t_r\epsilon_{sab}{p_a}{q_b}\epsilon_{ngr}C_{jsvg}C_{pdwn}b_{w}\sum^{\infty}_{q=0}\sum^{2q+1}_{m=0} \\[0.1in] 
& & \Re\left(S^{qm}_{vpd}\int^{s_2}_{s_1}\int^{r_2}_{r_1}\int^{y_2}_{y_1}\frac{\left({\bm R}\cdot{\bm e}_{12}\right)^{m}\left({\bm R}\cdot{\bm e}_3\right)^{2q+1-m}}{R^{2q+3}}(a_{0}+a_{1}r+a_{2}s+a_{3}rs)\mathrm{d}y\mathrm{d}r\mathrm{d}s\right)
\end{array}
\label{eq:tractionForceAniso}
\end{eqnarray}

The unknown parts of Eq.~\eqref{eq:tractionForceAniso} are the following series of integrals: 
\begin{eqnarray*}
K_{ijk} = \int^{s_2}_{s_1}\int^{r_2}_{r_1}\int^{y_2}_{y_1} \frac{\left({\bm R}\cdot{\bm e}_{12}\right)^{i}\left({\bm R}\cdot{\bm e}_3\right)^{j}}{R^{k}}\mathrm{d}y\mathrm{d}r\mathrm{d}s \\
K^r_{ijk} = \int^{s_2}_{s_1}\int^{r_2}_{r_1}\int^{y_2}_{y_1} \frac{\left({\bm R}\cdot{\bm e}_{12}\right)^{i}\left({\bm R}\cdot{\bm e}_3\right)^{j}}{R^{k}}r\mathrm{d}y\mathrm{d}r\mathrm{d}s \\
K^s_{ijk} = \int^{s_2}_{s_1}\int^{r_2}_{r_1}\int^{y_2}_{y_1} \frac{\left({\bm R}\cdot{\bm e}_{12}\right)^{i}\left({\bm R}\cdot{\bm e}_3\right)^{j}}{R^{k}}s\mathrm{d}y\mathrm{d}r\mathrm{d}s \\
K^{rs}_{ijk} = \int^{s_2}_{s_1}\int^{r_2}_{r_1}\int^{y_2}_{y_1} \frac{\left({\bm R}\cdot{\bm e}_{12}\right)^{i}\left({\bm R}\cdot{\bm e}_3\right)^{j}}{R^{k}}rs\mathrm{d}y\mathrm{d}r\mathrm{d}s,
\end{eqnarray*}
where $i+j=k-2$, and the common part of the integrands will hereafter be referred as $I_{ijk}$, i.e.
\begin{equation*}
I_{ijk} = \frac{\left({\bm R}\cdot{\bm e}_{12}\right)^{i}\left({\bm R}\cdot{\bm e}_3\right)^{j}}{R^{k}}.
\end{equation*}

\begin{figure}
\centering
\includegraphics[width=\linewidth]{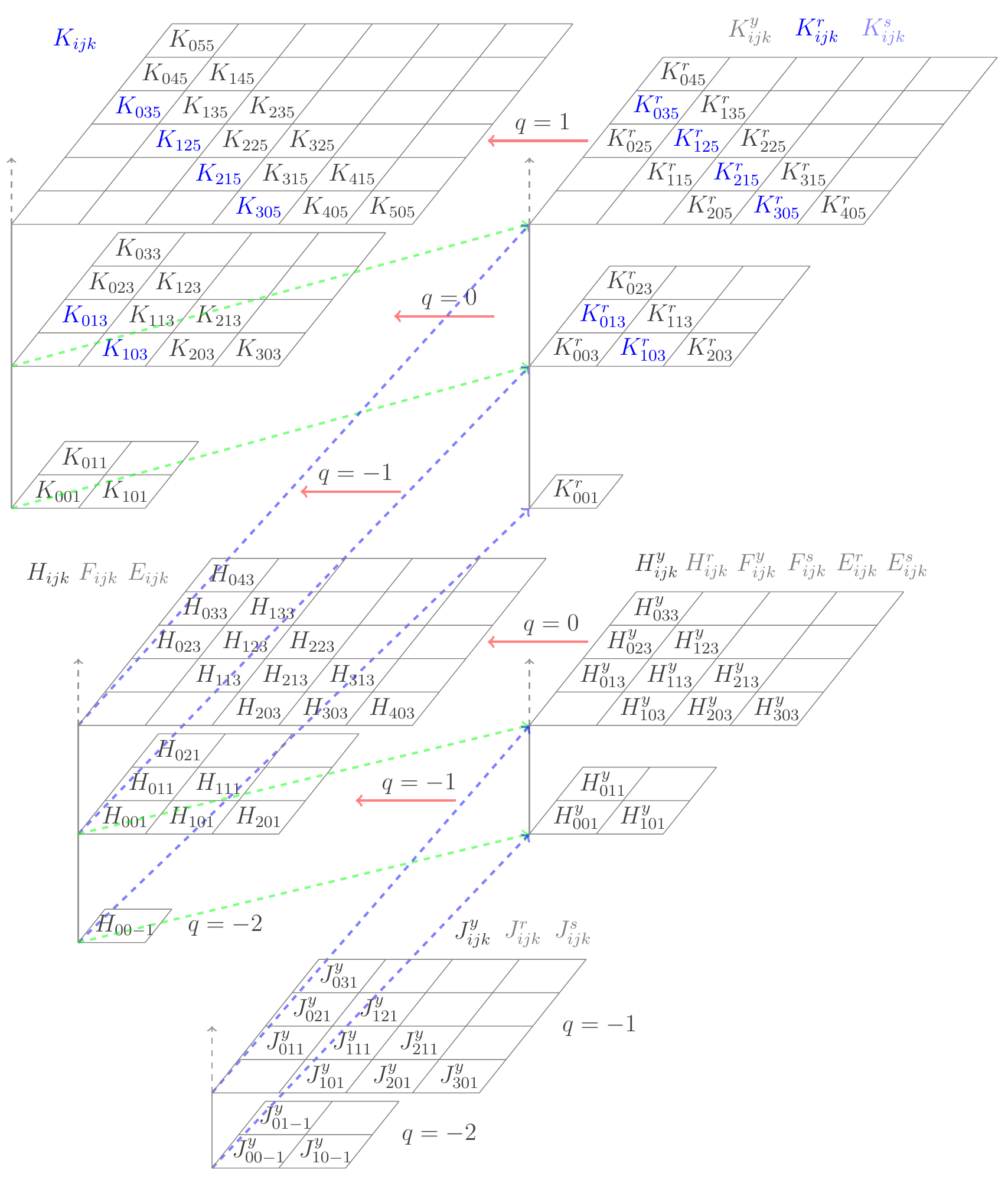}
\caption{Interface traction integration through hierarchical recurrence relations}
\label{fig:grid_triple}
\end{figure}

The main effort of this work lies in solving these series of triple integrals. We find that with the analytical solutions of a few seed integrals, all other integrals in the spherical harmonics expansion can be calculated analytically through recurrence relations. In Fig.~\ref{fig:grid_triple}, the required triple integrals for the traction force calculation are indicated with a blue color, and the rest triple integrals are necessary for the recursive integration to reach the required triple integrals of the same expansion order $q$ and the next expansion order $q+1$. As illustrated by the blue dashed lines in Fig.~\ref{fig:grid_triple}, the double integrals at the intermediate levels are needed for the calculations of the triple integrals at the highest levels, and the single integrals at the lowest levels must be used to calculate the double integrals at the intermediate levels.

The required recurrence relations are constructed using the following partial derivatives:
{\allowdisplaybreaks
\begin{eqnarray}
\partial_{y} I_{ijk} & = & i\alpha I_{(i-1)jk} + j\phi I_{i(j-1)k} - k \left({\bm R}\cdot{\bf t}\right) I_{ij(k-2)} \label{eq:partial1y}
\\
\partial_{r} I_{ijk} & = & i\beta I_{(i-1)jk} + j\theta I_{i(j-1)k} - k \left({\bm R}\cdot{\bf p}\right) I_{ij(k-2)} \label{eq:partial1r}
\\
\partial_{s} I_{ijk} & = & i\gamma I_{(i-1)jk} + j\psi I_{i(j-1)k} - k \left({\bm R}\cdot{\bf q}\right) I_{ij(k-2)} \label{eq:partial1s}
\\
\partial_{y} \left (y I_{ijk} \right) & = & I_{ijk} + y \, \partial_{y} I_{ijk} \label{eq:partial2y}
\\
\partial_{r} \left( r I_{ijk} \right) & = & I_{ijk} + r \, \partial_{r} I_{ijk} \label{eq:partial2r}
\\
\partial_{s} \left( s I_{ijk} \right) & = & I_{ijk} + s \, \partial_{s} I_{ijk} \label{eq:partial2s}
\\
\partial_{y} \left( {\bm R}\cdot{\bf t} I_{ijk} \right) & = & I_{ijk} + {\bm R}\cdot{\bf t} \, \partial_{y} I_{ijk} \label{eq:partial3y}
\\
\partial_{r} \left( {\bm R}\cdot{\bf p} I_{ijk} \right) & = & I_{ijk} + {\bm R}\cdot{\bf p} \, \partial_{r} I_{ijk} \label{eq:partial3r}
\\
\partial_{s} \left( {\bm R}\cdot{\bf q} I_{ijk} \right) & = & I_{ijk} + {\bm R}\cdot{\bf q} \, \partial_{s} I_{ijk} \label{eq:partial3s}
\end{eqnarray}
}
where $\alpha = {\bm t}\cdot{\bm e}_{12}$, $\beta = {\bm p}\cdot{\bm e}_{12}$, $\gamma = {\bm q}\cdot{\bm e}_{12}$, $\phi = {\bm t}\cdot{\bm e}_{3}$, $\theta = {\bm p}\cdot{\bm e}_{3}$, and $\psi = {\bm q}\cdot{\bm e}_{3}$.

The first three recurrence relations Eq.~\eqref{eq:Kyrs1}, Eq.~\eqref{eq:Kyrs2}, and Eq.~\eqref{eq:Kyrs3} for the triple integrals are obtained by successive integrations over $y$, $r$, and $s$ of the partial derivatives Eq.~\eqref{eq:partial1y}, Eq.~\eqref{eq:partial1r}, and Eq.~\eqref{eq:partial1s}, respectively. Another two recurrence relations Eq.~\eqref{eq:K1} and Eq.~\eqref{eq:K2} can be verified by the triple integral definitions.
\begin{eqnarray}
K^y_{ij\left(k+2\right)} + cK^r_{ij\left(k+2\right)} + dK^s_{ij\left(k+2\right)} & = & \frac{1}{k} \left[ i\alpha K_{\left(i-1\right)jk} + j\phi K_{i\left(j-1\right)k} - E_{ijk} \right] \label{eq:Kyrs1}
\\
cK^y_{ij\left(k+2\right)} + K^r_{ij\left(k+2\right)} + fK^s_{ij\left(k+2\right)} & = & \frac{1}{k} \left[ i\beta K_{\left(i-1\right)jk} + j\theta K_{i\left(j-1\right)k} - F_{ijk} \right] \label{eq:Kyrs2}
\\
dK^y_{ij\left(k+2\right)} + fK^r_{ij\left(k+2\right)} + K^s_{ij\left(k+2\right)} & = & \frac{1}{k} \left[ i\gamma K_{\left(i-1\right)jk} + j\psi K_{i\left(j-1\right)k} - H_{ijk} \right] \label{eq:Kyrs3}
\\
K_{\left(i+1\right)jk} & = & \alpha K^y_{ijk} + \beta K^r_{ijk} + \gamma K^s_{ijk} \label{eq:K1}
\\
K_{i\left(j+1\right)k} & = & \phi K^y_{ijk} + \theta K^r_{ijk} + \psi K^s_{ijk} \label{eq:K2},
\end{eqnarray}
where $c = {\bm p}\cdot{\bm t}$, $d = {\bm q}\cdot{\bm t}$, $f = {\bm p}\cdot{\bm q}$, $K^y_{ijk} = \int^{s_2}_{s_1}\int^{r_2}_{r_1}\int^{y_2}_{y_1} I_{ijk} y\mathrm{d}y\mathrm{d}r\mathrm{d}s$, $E_{ijk} = \int^{s_2}_{s_1}\int^{r_2}_{r_1} I_{ijk} \mathrm{d}r\mathrm{d}s$, $F_{ijk} = \int^{s_2}_{s_1}\int^{y_2}_{y_1} I_{ijk} \mathrm{d}y\mathrm{d}s$, and $H_{ijk} = \int^{r_2}_{r_1}\int^{y_2}_{y_1} I_{ijk} \mathrm{d}y\mathrm{d}r$. The double integrals $H_{ijk}$, $F_{ijk}$, and $E_{ijk}$ need to be previously calculated using the second set of recurrence relations given below.

The first two recurrence relations Eq.~\eqref{eq:Hyr1} and Eq.~\eqref{eq:Hyr2} for the double integrals $H_{ijk}$ are obtained by successive integrations over $y$ and $r$ of the partial derivatives Eq.~\eqref{eq:partial1y} and Eq.~\eqref{eq:partial1r}, respectively. The third recurrence relation Eq.~\eqref{eq:Hyr3} for the double integrals $H_{ijk}$ is obtained by successive integrations over $y$ and $r$ of the partial derivatives Eq.~\eqref{eq:partial2y} and Eq.~\eqref{eq:partial2r}, and then summation of these two integration equations. Another two recurrence relations Eq.~\eqref{eq:H1} and Eq.~\eqref{eq:H2} can be verified by the double integral definitions. Analogously, the recurrence relations for the double integrals $F_{ijk}$ Eq.~\eqref{eq:Fys1} to Eq.~\eqref{eq:F2} and for the double integrals $E_{ijk}$ Eq.~\eqref{eq:Ers1} to Eq.~\eqref{eq:E2} are obtained using the corresponding partial derivatives and double integral definitions.
{\allowdisplaybreaks
\begin{eqnarray}
H^y_{ij\left(k+2\right)} + cH^r_{ij\left(k+2\right)} + dsH_{ij\left(k+2\right)} = \frac{1}{k} \left[ i\alpha H_{\left(i-1\right)jk} + j\phi H_{i\left(j-1\right)k} - J^r_{ijk} \right] \label{eq:Hyr1}
\\
cH^y_{ij\left(k+2\right)} + H^r_{ij\left(k+2\right)} + fsH_{ij\left(k+2\right)} = \frac{1}{k} \left[ i\beta H_{\left(i-1\right)jk} + j\theta H_{i\left(j-1\right)k} - J^y_{ijk} \right] \label{eq:Hyr2}
\\
\begin{array}{r}
dsH^y_{ij\left(k+2\right)} + fsH^r_{ij\left(k+2\right)} + s^2 H_{ij\left(k+2\right)} = \frac{1}{k} \left[ i\gamma s H_{\left(i-1\right)jk} + j\psi s H_{i\left(j-1\right)k} \right. \\ 
\left. + \left(k-i-j-2\right)H_{ijk} + rJ^y_{ijk} +yJ^r_{ijk} \right]
\end{array} \label{eq:Hyr3}
\\
H_{\left(i+1\right)jk} = \alpha H^y_{ijk} + \beta H^r_{ijk} + \gamma s  H_{ijk} \label{eq:H1}
\\
H_{i\left(j+1\right)k} = \phi H^y_{ijk} + \theta H^r_{ijk} + \psi s  H_{ijk} \label{eq:H2}
\\
F^y_{ij\left(k+2\right)} + dF^s_{ij\left(k+2\right)} + crF_{ij\left(k+2\right)} = \frac{1}{k} \left[ i\alpha F_{\left(i-1\right)jk} + j\phi F_{i\left(j-1\right)k} - J^s_{ijk} \right] \label{eq:Fys1}
\\
dF^y_{ij\left(k+2\right)} + F^s_{ij\left(k+2\right)} + frF_{ij\left(k+2\right)} = \frac{1}{k} \left[ i\gamma F_{\left(i-1\right)jk} + j\psi F_{i\left(j-1\right)k} - J^y_{ijk} \right] \label{eq:Fys2}
\\
\begin{array}{r}
crF^y_{ij\left(k+2\right)} + frF^s_{ij\left(k+2\right)} + r^2 F_{ij\left(k+2\right)} = \frac{1}{k} \left[ i\beta r F_{\left(i-1\right)jk} + j\theta r F_{i\left(j-1\right)k} \right. \\
\left. + \left(k-i-j-2\right)F_{ijk} + yJ^s_{ijk} +sJ^y_{ijk} \right]
\end{array} \label{eq:Fys3}
\\
F_{\left(i+1\right)jk} = \alpha F^y_{ijk} + \beta r F_{ijk} + \gamma F^s_{ijk} \label{eq:F1}
\\
F_{i\left(j+1\right)k} = \phi F^y_{ijk} + \theta r F_{ijk} + \psi F^s_{ijk} \label{eq:F2}
\\
E^r_{ij\left(k+2\right)} + fE^s_{ij\left(k+2\right)} + cyE_{ij\left(k+2\right)} = \frac{1}{k} \left[ i\beta E_{\left(i-1\right)jk} + j\theta E_{i\left(j-1\right)k} - J^s_{ijk} \right] \label{eq:Ers1}
\\
fE^r_{ij\left(k+2\right)} + E^s_{ij\left(k+2\right)} + dyE_{ij\left(k+2\right)} = \frac{1}{k} \left[ i\gamma E_{\left(i-1\right)jk} + j\psi E_{i\left(j-1\right)k} - J^r_{ijk} \right] \label{eq:Ers2}
\\
\begin{array}{r}
cyE^r_{ij\left(k+2\right)} + dyE^r_{ij\left(k+2\right)} + y^2 E_{ij\left(k+2\right)}  =  \frac{1}{k} \left[ i\alpha y E_{\left(i-1\right)jk} + j\phi y E_{i\left(j-1\right)k} \right. \\
\left. + \left(k-i-j-2\right)E_{ijk} + sJ^r_{ijk} +rJ^s_{ijk} \right]
\end{array} \label{eq:Ers3}
\\
E_{\left(i+1\right)jk} = \alpha y E_{ijk} + \beta E^r_{ijk} + \gamma E^s_{ijk} \label{eq:E1}
\\
E_{i\left(j+1\right)k} = \phi y E_{ijk} + \theta E^r_{ijk} + \psi E^s_{ijk} \label{eq:E2},
\end{eqnarray}
}
where $H^{y}_{ijk} = \int^{r_2}_{r_1}\int^{y_2}_{y_1} I_{ijk} y\mathrm{d}y\mathrm{d}r$, $H^{r}_{ijk} = \int^{r_2}_{r_1}\int^{y_2}_{y_1} I_{ijk} r\mathrm{d}y\mathrm{d}r$, $F^{y}_{ijk} = \int^{s_2}_{s_1}\int^{y_2}_{y_1} I_{ijk} y\mathrm{d}y\mathrm{d}s$, $F^{s}_{ijk} = \int^{s_2}_{s_1}\int^{y_2}_{y_1} I_{ijk} s\mathrm{d}y\mathrm{d}s$, $E^r_{ijk} = \int^{s_2}_{s_1}\int^{r_2}_{r_1} I_{ijk} r\mathrm{d}r\mathrm{d}s$, $E^s_{ijk} = \int^{s_2}_{s_1}\int^{r_2}_{r_1} I_{ijk} s\mathrm{d}r\mathrm{d}s$, $J^{y}_{ijk} = \int^{y_2}_{y_1} I_{ijk} \mathrm{d}y$, $J^{r}_{ijk} = \int^{r_2}_{r_1} I_{ijk} \mathrm{d}r$, and $J^{s}_{ijk} = \int^{s_2}_{s_1} I_{ijk} \mathrm{d}s$. The single integrals $J^{y}_{ijk}$, $J^{r}_{ijk}$, and $J^{s}_{ijk}$ must be formerly calculated using the third set of recurrence relations given below.

The first two recurrence relations Eq.~\eqref{eq:Jy1} and Eq.~\eqref{eq:Jy2} for the single integrals $J^{y}_{ijk}$ are obtained by integration over $y$ of the partial derivative Eq.~\eqref{eq:partial1y} and multiplying both sides of the integration equation by $\alpha$ and $\phi$, respectively. The third recurrence relation for the single integrals $J^{y}_{ijk}$ Eq.~\eqref{eq:Jy3} is obtained by integration over $y$ of the partial derivative Eq.~\eqref{eq:partial3y}. Analogously, the recurrence relations for the single integrals $J^{r}_{ijk}$ (from Eq.~\eqref{eq:Jr1} to Eq.~\eqref{eq:Jr3}) and $J^{s}_{ijk}$ (from Eq.~\eqref{eq:Js1} to Eq.~\eqref{eq:Js3}) are obtained using the corresponding partial derivatives.
{\allowdisplaybreaks
\begin{eqnarray}
J^{y}_{\left(i+1\right)j\left(k+2\right)}  = \frac{\alpha}{k} \left[ i\alpha J^{y}_{\left(i-1\right)jk} + j\phi J^{y}_{i\left(j-1\right)k} - I_{ijk}  \right]  + J^{y}_{ij\left(k+2\right)}\left[\left(\beta - \alpha c\right)r + \left(\gamma - \alpha d\right)s\right] \label{eq:Jy1}
\\
J^{y}_{i\left(j+1\right)\left(k+2\right)}  = \frac{\phi}{k} \left[ i\alpha J^{y}_{\left(i-1\right)jk} + j\phi J^{y}_{i\left(j-1\right)k} - I_{ijk}  \right]  + J^{y}_{ij\left(k+2\right)}\left[\left(\theta - \phi c\right)r + \left(\psi - \phi d\right)s\right] \label{eq:Jy2}
\\
\begin{array}{r}
J^{y}_{ij\left(k+2\right)} = \frac{1}{k\left[R^2-\left({\bm R}\cdot{\bf t}\right)^2\right]} \left[ {\bm R}\cdot{\bf t} I_{ijk} + i\left[\left(\beta - \alpha c\right)r + \left(\gamma - \alpha d\right)s\right]J^{y}_{\left(i-1\right)jk}  \right. \\ 
\left. + j\left[\left(\theta - \phi c\right)r + \left(\psi - \phi d\right)s\right]J^{y}_{i\left(j-1\right)k} + \left(k-1-i-j\right)J^{y}_{ijk} \right]
\end{array} \label{eq:Jy3}
\\
J^{r}_{\left(i+1\right)j\left(k+2\right)}  = \frac{\beta}{k} \left[ i\beta J^{r}_{\left(i-1\right)jk} + j\theta J^{r}_{i\left(j-1\right)k} - I_{ijk}  \right]  + J^{r}_{ij\left(k+2\right)}\left[\left(\alpha - \beta c\right)y + \left(\gamma - \beta f\right)s\right] \label{eq:Jr1}
\\
J^{r}_{i\left(j+1\right)\left(k+2\right)}  = \frac{\theta}{k} \left[ i\beta J^{r}_{\left(i-1\right)jk} + j\theta J^{r}_{i\left(j-1\right)k} - I_{ijk}  \right]  + J^{r}_{ij\left(k+2\right)}\left[\left(\phi - \theta c\right)y + \left(\psi - \theta f\right)s\right] \label{eq:Jr2}
\\
\begin{array}{r}
J^{r}_{ij\left(k+2\right)} = \frac{1}{k\left[R^2-\left({\bm R}\cdot{\bf p}\right)^2\right]} \left[ {\bm R}\cdot{\bf p} I_{ijk} + i\left[\left(\alpha - \beta c\right)y + \left(\gamma - \beta f\right)s\right]J^{r}_{\left(i-1\right)jk}  \right. \\
\left. + j\left[\left(\phi - \theta c\right)y + \left(\psi - \theta f\right)s\right]J^{r}_{i\left(j-1\right)k} + \left(k-1-i-j\right)J^{r}_{ijk} \right]
\end{array} \label{eq:Jr3}
\\
J^{s}_{\left(i+1\right)j\left(k+2\right)}  = \frac{\gamma}{k} \left[ i\gamma J^{s}_{\left(i-1\right)jk} + j\psi J^{s}_{i\left(j-1\right)k} - I_{ijk}  \right]  + J^{s}_{ij\left(k+2\right)}\left[\left(\alpha - \gamma d\right)y + \left(\beta - \gamma f\right)r\right] \label{eq:Js1}
\\
J^{s}_{i\left(j+1\right)\left(k+2\right)}  = \frac{\psi}{k} \left[ i\gamma J^{s}_{\left(i-1\right)jk} + j\psi J^{s}_{i\left(j-1\right)k} - I_{ijk}  \right]  + J^{s}_{ij\left(k+2\right)}\left[\left(\phi - \psi d\right)y + \left(\theta - \psi f\right)r\right] \label{eq:Js2}
\\
\begin{array}{r}
J^{s}_{ij\left(k+2\right)} = \frac{1}{k\left[R^2-\left({\bm R}\cdot{\bf q}\right)^2\right]} \left[ {\bm R}\cdot{\bf q} I_{ijk} + i\left[\left(\alpha - \gamma d\right)y + \left(\beta - \gamma f\right)r\right]J^{s}_{\left(i-1\right)jk} \right. \\
\left. + j\left[\left(\phi - \psi d\right)y + \left(\theta - \psi f\right)r\right]J^{s}_{i\left(j-1\right)k} + \left(k-1-i-j\right)J^{s}_{ijk} \right]
\end{array} \label{eq:Js3}
\end{eqnarray}
}

As illustrated in Fig.~\ref{fig:grid_triple}, a number of seed integrals have to be first calculated before starting the recursive integrations. The seed single integrals $J^{y}_{00-1}$, $J^{r}_{00-1}$, and $J^{s}_{00-1}$ can be directly calculated using the corresponding analytical solutions Eq.~\eqref{eq:Jy00m1}, Eq.~\eqref{eq:Jr00m1}, and Eq.~\eqref{eq:Js00m1}, respectively.
\begin{eqnarray}
J^{y}_{00-1} & = &\frac{1}{2} \left\{ \left[R^2-\left({\bm R}\cdot{\bf t}\right)^2\right] \ln\left(R+{\bm R}\cdot{\bf t}\right) + {\bm R}\cdot{\bf t} R \right\} \label{eq:Jy00m1}
\\
J^{r}_{00-1} & = & \frac{1}{2} \left\{ \left[R^2-\left({\bm R}\cdot{\bf p}\right)^2\right] \ln\left(R+{\bm R}\cdot{\bf p}\right) + {\bm R}\cdot{\bf p} R \right\} \label{eq:Jr00m1}
\\
J^{s}_{00-1} & = & \frac{1}{2} \left\{ \left[R^2-\left({\bm R}\cdot{\bf q}\right)^2\right] \ln\left(R+{\bm R}\cdot{\bf q}\right) + {\bm R}\cdot{\bf q} R \right\} \label{eq:Js00m1}
\end{eqnarray}
The rest single integrals are all calculated using the recurrence relations Eq.~\eqref{eq:Jy1} to Eq.~\eqref{eq:Js3}. For each type of single integrals, the three recurrence relations are used to increase $i$, $j$, and $k$ indices, respectively, e.g. $J^{y}_{00-1} \rightarrow  J^{y}_{10-1}$, $J^{y}_{00-1} \rightarrow  J^{y}_{01-1}$, and $J^{y}_{00-1} \rightarrow  J^{y}_{001}$.

The seed double integrals $H_{001}$, $H_{00-1}$, $F_{001}$, $F_{00-1}$, $E_{001}$, and $E_{00-1}$ are calculated using the analytical solution of $H_{003}$, $F_{003}$, and $E_{003}$ given in Eq.~\eqref{eq:H003}, Eq.~\eqref{eq:F003}, and Eq.~\eqref{eq:E003} and applying inverse recurrence relations Eq.~\eqref{eq:H00k}, Eq.~\eqref{eq:F00k}, and Eq.~\eqref{eq:E00k}. These inverse recurrence relations are obtained by solving the first three recurrence relations for each type of the double integrals, i.e. Eq.~\eqref{eq:Hyr1} to Eq.~\eqref{eq:Hyr3} for $H_{ijk}$, Eq.~\eqref{eq:Fys1} to Eq.~\eqref{eq:Fys3} for $F_{ijk}$, and Eq.~\eqref{eq:Ers1} to Eq.~\eqref{eq:Ers3} for $E_{ijk}$.
{\allowdisplaybreaks
\begin{eqnarray}
H_{003} = \frac{2}{\sqrt{s^2(1+2cdf-c^2-d^2-f^2)}}\arctan\left[\frac{(1-c)(R+y-r)+(d-f)s}{\sqrt{s^2(1+2cdf-c^2-d^2-f^2)}}\right] \label{eq:H003}
\\
F_{003} = \frac{2}{\sqrt{r^2(1+2cdf-c^2-d^2-f^2)}}\arctan\left[\frac{(1-d)(R+y-s)+(c-f)r}{\sqrt{r^2(1+2cdf-c^2-d^2-f^2)}}\right] \label{eq:F003}
\\
E_{003} = \frac{2}{\sqrt{y^2(1+2cdf-c^2-d^2-f^2)}}\arctan\left[\frac{(1-f)(R+r-s)+(c-d)y}{\sqrt{y^2(1+2cdf-c^2-d^2-f^2)}}\right] \label{eq:E003}
\\
\begin{array}{l l l}
H_{00k} & = & \frac{1}{(k-2)(1-c^2)} \left\{ k[(1-c^2)(1-d^2)-(f-cd)^2]s^2 H_{00(k+2)} \right. \\ 
& & \left. - [(1-c^2)r+(f-cd)s]J^{y}_{00k} - [(1-c^2)y+(d-cf)s]J^{r}_{00k} \right\}
\end{array} \label{eq:H00k}
\\
\begin{array}{l l l}
F_{00k} & = & \frac{1}{(k-2)(1-d^2)} \left\{ k[(1-c^2)(1-d^2)-(f-cd)^2]r^2 F_{00(k+2)} \right. \\ 
& & \left. - [(1-d^2)s+(f-cd)r]J^{y}_{00k} - [(1-d^2)y+(c-df)r]J^{s}_{00k} \right\}
\end{array} \label{eq:F00k}
\\
\begin{array}{l l l}
E_{00k} & = & \frac{1}{(k-2)(1-f^2)} \left\{ k[(1-c^2)(1-f^2)-(d-cf)^2]y^2 E_{00(k+2)} \right. \\ 
&  & \left. - [(1-f^2)s+(d-cf)y]J^{r}_{00k} - [(1-f^2)r+(c-df)y]J^{s}_{00k} \right\}
\end{array} \label{eq:E00k}
\end{eqnarray}
}
The other double integrals are all calculated using the recurrence relations Eq.~\eqref{eq:Hyr1} to Eq.~\eqref{eq:E2}. The first three recurrences Eq.~\eqref{eq:Hyr1} to Eq.~\eqref{eq:Hyr3} are used to calculate the double integrals $H^{y}_{ijk+2}$ and $H^{r}_{ijk+2}$ from the available double integrals $H_{ijk}$ and single integrals $J^{y}_{ijk}$ and $J^{r}_{ijk}$ at a lower $k$ index, see Fig.~\ref{fig:grid_triple}. The last two recurrence relations Eq.~\eqref{eq:H1} and Eq.~\eqref{eq:H2} are used to calculate the double integrals $H_{ijk}$ at higher $i$ and $j$ indices, respectively.

The seed triple integral $K_{001}$ can be directly calculated from the seed double integrals $H_{001}$, $F_{001}$, and $E_{001}$ using Eq.~\eqref{eq:K001}, which is obtained by successive integrations over $y$, $r$, and $s$ of the partial derivatives Eq.~\eqref{eq:partial2y}, Eq.~\eqref{eq:partial2r}, and Eq.~\eqref{eq:partial2s}, and then summation of these three integration equations.
\begin{eqnarray}
\begin{array}{c}
\left(3+i+j-k\right)K_{ijk} = yE_{ijk} + rF_{ijk} + sH_{ijk}
\\
K_{001} = \frac{1}{2} \left(yE_{001} + rF_{001} + sH_{001}\right)
\end{array}
\label{eq:K001}
\end{eqnarray}
The remaining triple integrals are all calculated using the recurrence relations Eq.~\eqref{eq:Kyrs1} to Eq.~\eqref{eq:K2}. The first three recurrences Eq.~\eqref{eq:Kyrs1} to Eq.~\eqref{eq:Kyrs3} are used to calculated the double integrals $K^{y}_{ijk+2}$, $K^{r}_{ijk+2}$, and $K^{s}_{ijk+2}$ from the available triple integrals $K_{ijk}$ and the double integrals $H_{ijk}$, $F_{ijk}$, and $E_{ijk}$ at a lower $k$ index, see Fig.~\ref{fig:grid_triple}. The last two recurrence relations Eq.~\eqref{eq:K1} and Eq.~\eqref{eq:K2} are used to calculate the triple integrals $K_{ijk}$ at higher $i$ and $j$ indices, respectively.

The triple integrals $K^{rs}_{ijk}$ can be directly calculated from $K_{ijk}$, $K^{r}_{ijk}$, $H^{r}_{ijk}$, $F_{ijk}$, and $E^{r}_{ijk}$ using the relations Eq.~\eqref{eq:Krs1} to Eq.~\eqref{eq:Krs3}. The first equation is obtained by successive integrations over $y$, $r$, and $s$ of the partial derivative Eq.~\eqref{eq:partial2r}. The last two equations are obtained by multiplying the partial differential equations Eq.~\eqref{eq:partial1y} and Eq.~\eqref{eq:partial1s} by $r$ on both sides, and successive integrations over $y$, $r$, and $s$ of these two partial differential equations, respectively.
\begin{eqnarray}
K^{rr}_{ij\left(k+2\right)} + cK^{yr}_{ij\left(k+2\right)} + fK^{rs}_{ij\left(k+2\right)} & = & \frac{1}{k} \left[K_{ijk} + i\beta K^r_{\left(i-1\right)jk} + j\theta K^r_{i\left(j-1\right)k} - rF_{ijk} \right] \label{eq:Krs1}
\\
cK^{rr}_{ij\left(k+2\right)} + K^{yr}_{ij\left(k+2\right)} + dK^{rs}_{ij\left(k+2\right)} & = & \frac{1}{k} \left[ i\alpha K^r_{\left(i-1\right)jk} + j\phi K^r_{i\left(j-1\right)k} - E^r_{ijk} \right] \label{eq:Krs2}
\\
fK^{rr}_{ij\left(k+2\right)} + dK^{yr}_{ij\left(k+2\right)} + K^{rs}_{ij\left(k+2\right)} & = & \frac{1}{k} \left[ i\gamma K^r_{\left(i-1\right)jk} + j\psi K^r_{i\left(j-1\right)k} - H^r_{ijk} \right] \label{eq:Krs3}
\end{eqnarray}

\section{Results}
To evaluate the accuracy and efficiency of our analytical traction force calculation, we perform $q_{max}$ (spherical harmonics expansion order) convergence tests, and compare with Gaussian quadrature numerical integrations.
 
\begin{figure}
\centering
\includegraphics[width=0.7\linewidth]{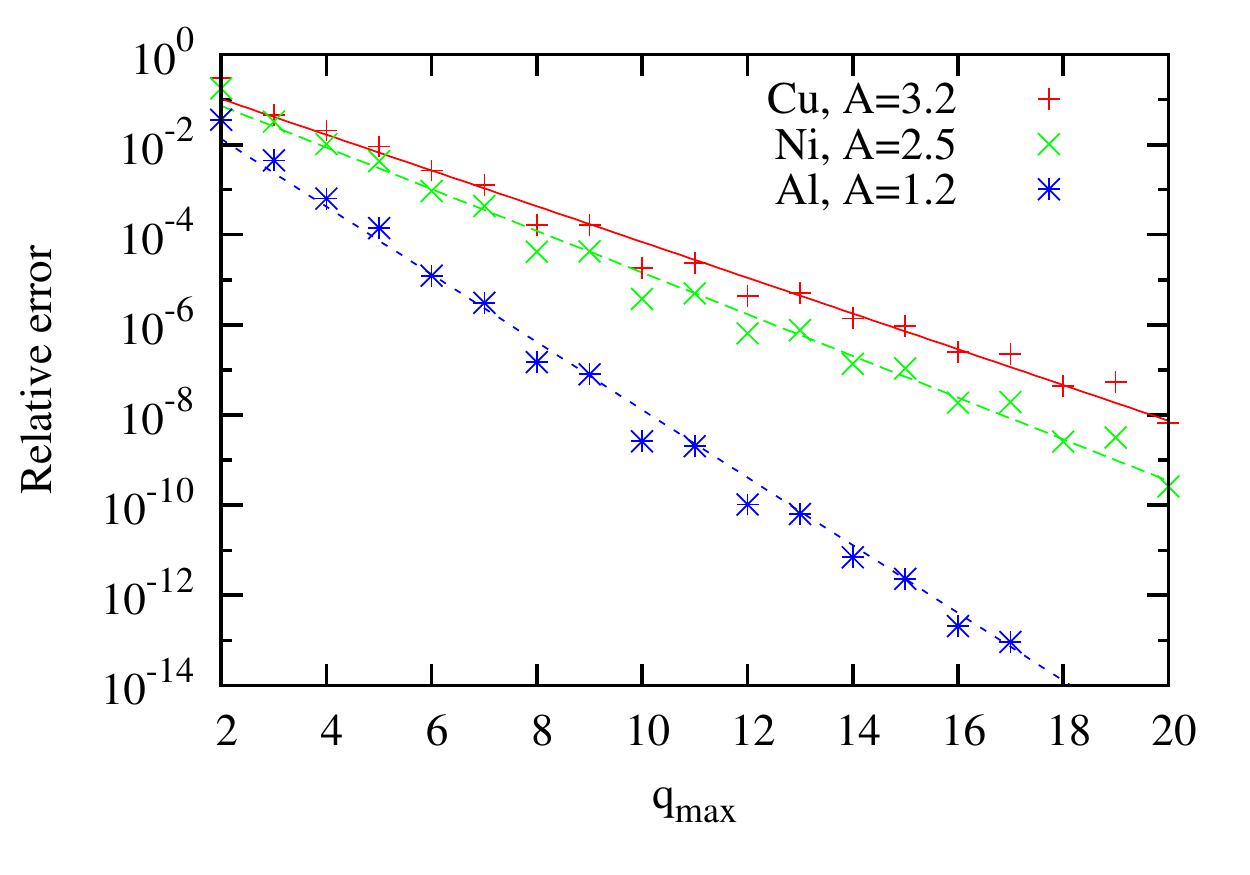}
\caption{Convergence of traction force calculation as the spherical harmonics expansion order $q_{max}$ increases for different materials in terms of the elastic anisotropy ratio $A$.}
\label{fig:QmaxError}
\end{figure}

The infinite series of spherical harmonics expansions must be truncated in practice. Figure~\ref{fig:QmaxError} shows how the relative error of the traction force calculation evolves as the spherical harmonics expansion order $q_{max}$ increases. Similar to the dislocation interaction force calculations of Aubry and Arsenlis~\cite{Aubry&Arsenlis2013}, the traction force calculations converge faster for materials with lower elastic anisotropy ratios, which means high orders of spherical harmonics expansion are needed for materials with high elastic anisotropy.

\begin{figure}
\centering
\includegraphics[width=0.7\linewidth]{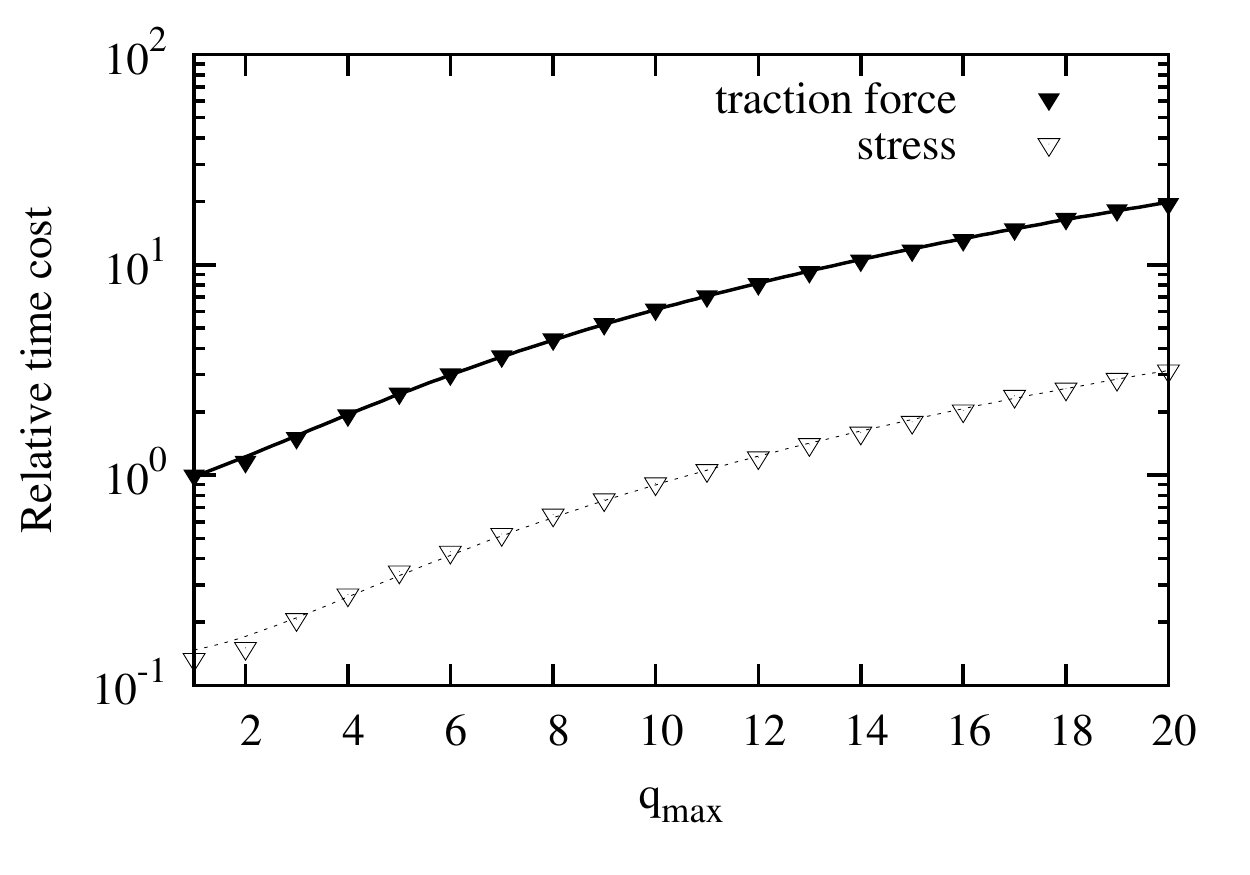}
\caption{Quadratic computation cost growth of interface traction force and dislocation stress field calculations with the spherical harmonics expansion order $q_{max}$.}
\label{fig:QmaxTime}
\end{figure}

How the computation cost increases with the order of spherical harmonics expansion is presented in Fig.~\ref{fig:QmaxTime}. The cost of our traction force calculation grows quadratically as the stress field calculation using the analytical expressions given in Ref.~\cite{Aubry&Arsenlis2013}. For the traction force calculation, using $q_{max} = 20$ is twenty times more expensive than using $q_{max} = 1$. The computation cost ratio between the traction force and stress field calculations is rather insensitive to the spherical harmonics expansion order, and changes from seven to six when $q_{max}$ increases from one to twenty.

We now compare our analytical interface traction force integrations with Gaussian quadrature numerical integrations using the analytical stress field expressions of Ref.~\cite{Aubry&Arsenlis2013}. As both the traction force and stress field calculations use the same spherical harmonics expansions of the derivatives of Green's function, the comparison of analytical and numerical integrations for a fixed $q_{max}$ is roughly the same when $q_{max}$ is changed from one to twenty.

\begin{figure}
\centering
\includegraphics[width=0.7\linewidth]{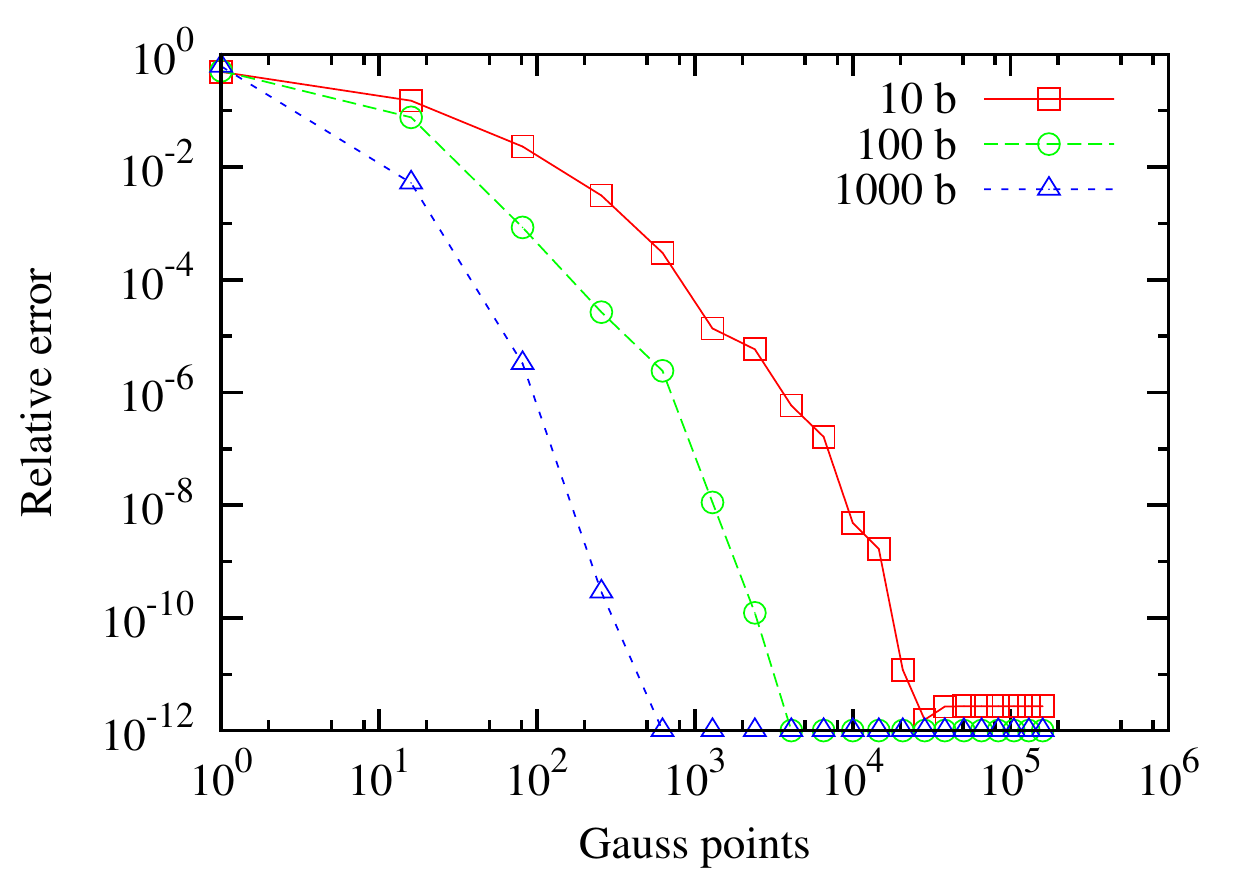}
\caption{Comparison of Gaussian quadrature numerical integration and analytical triple integration using recurrence relations for Ni with $q_{max}=10$ for different dislocation--interface distances in the unit of the Burgers vector's magnitude $b$.}
\label{fig:AnaNumError}
\end{figure}

Figure~\ref{fig:AnaNumError} depicts how the relative error between analytical and numerical integrations varies as the number of Gauss quadrature points increases in the numerical integrations. Similar to the traction force calculations of Queyreau et al.~\cite{Queyreau2014} for isotropic elastic media, the relative error decreases faster for larger dislocation--interface distances as the number of Gauss points increases in the numerical integrations. While the analytical solution is used as a reference to assess the error in numerical integrations, the comparison with numerical integrations using a large number of Gauss points can verify the correct implementation of the analytical traction force integration.

\begin{figure}
\centering
\includegraphics[width=0.7\linewidth]{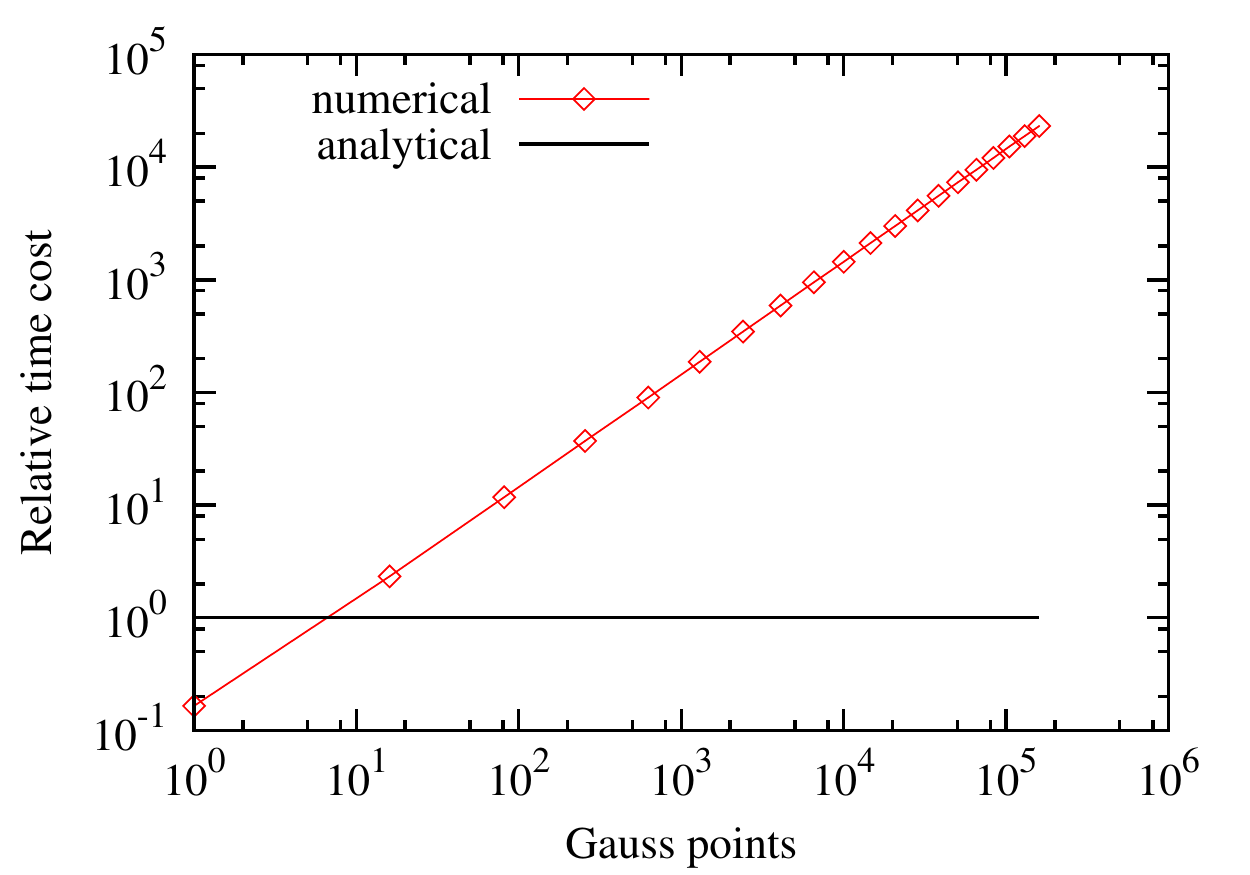}
\caption{Computation cost comparison of Gaussian quadrature numerical integration and analytical triple integration using recurrence relations for a fixed $q_{max}$ of 10.}
\label{fig:AnaNumTime}
\end{figure}

Figure~\ref{fig:AnaNumTime} illustrates the computation cost comparison of analytical and numerical traction force integrations. The analytical integration becomes more efficient when the number of Gauss points exceeds eight in the numerical integration. With eight Gauss points, the error of the numerical integration is above one percent for a large dislocation--interface distance of 1000~$b$, and above ten percent for smaller dislocation--interface distances of 100~$b$ and 10~$b$, Fig.~\ref{fig:AnaNumError}. Keep in mind that the traction force calculation is only for one dislocation segment and one interface element, and the computation error can escalate in DD--FE/BE simulations with larger numbers of dislocation segments and interface elements over many correlated time steps. As proposed by Weygand et al.~\cite{Weygand2002}, a minimum of one hundred integration points has to be used for numerical integrations of surface traction forces. As shown in Fig.~\ref{fig:AnaNumTime}, compared with a numerical integration using one hundred integration points, our analytical integration is more than one order of magnitude faster in speed.

\begin{figure}
\centering
\includegraphics[width=0.6\linewidth]{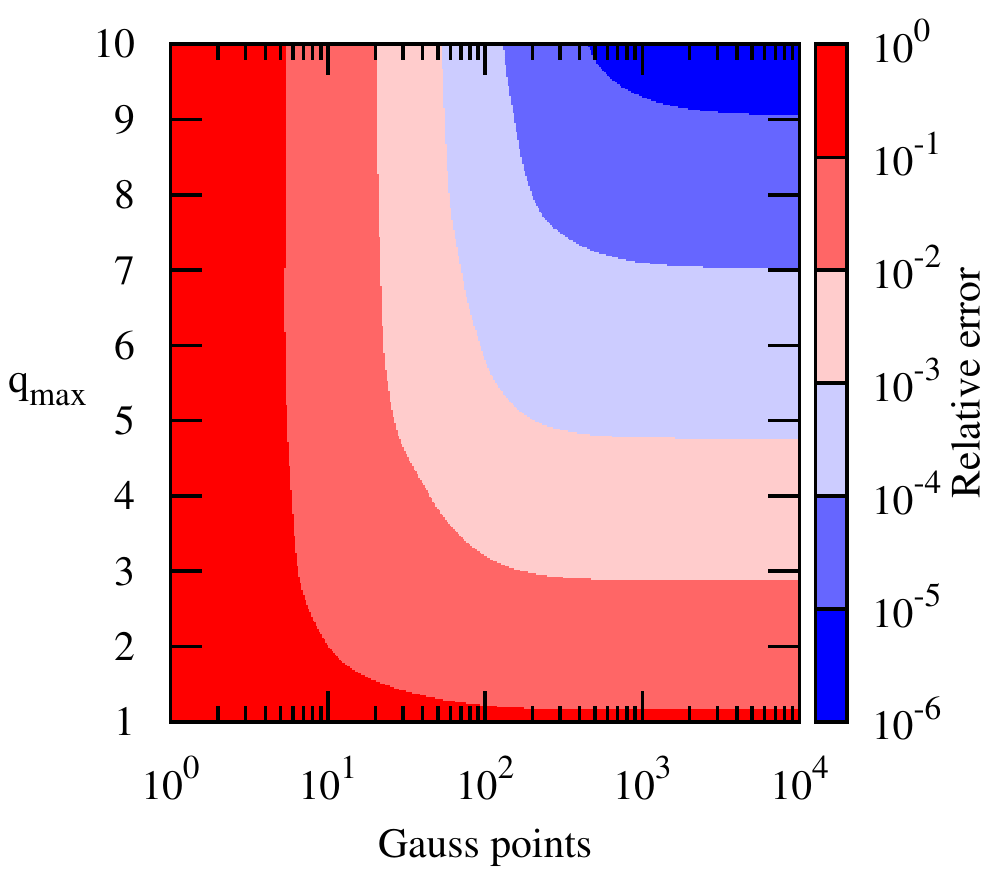}
\includegraphics[width=0.6\linewidth]{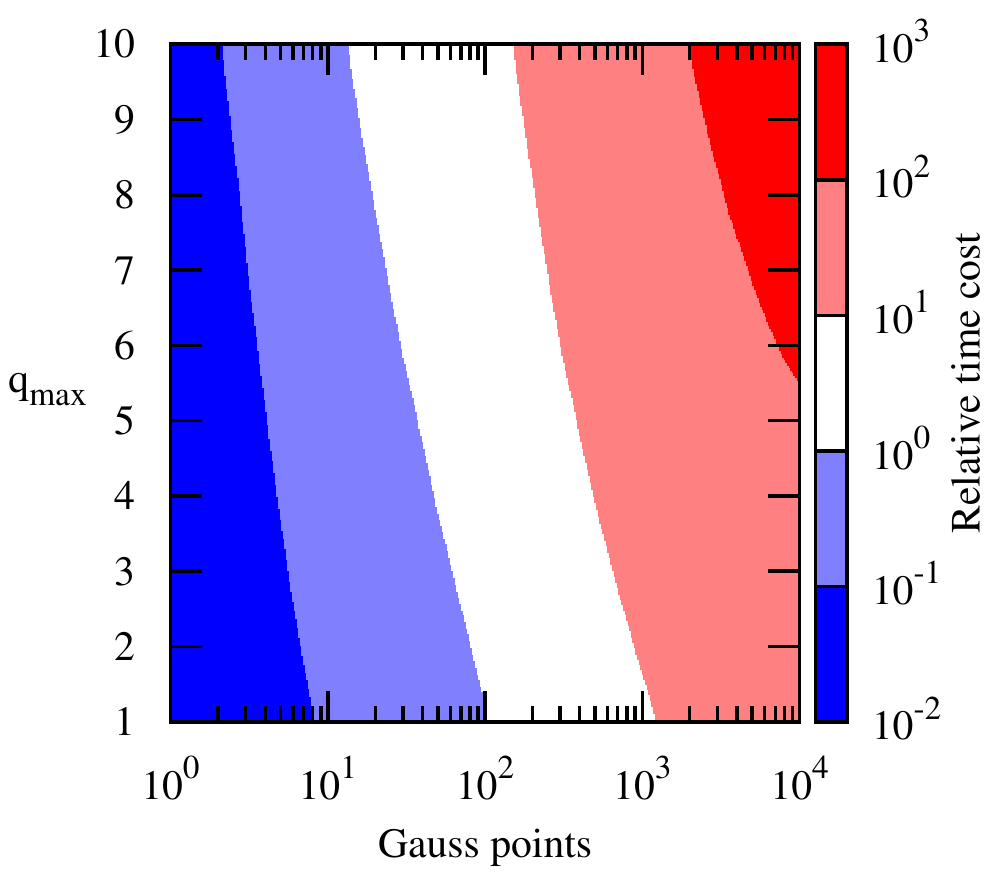}
\caption{Computation error and time cost of numerical integrations with respect to an analytical traction force integration using $q_{max} = 20$.}
\label{fig:NumErrorTimeContourAnaQmax20}
\end{figure}

Despite the obvious advantages of the analytical interface traction force integration, the algorithm implementation into a specific DD-FE/BE code can bring a considerable overhead to a short research project. In such a case, using simpler analytical stress field expressions and standard Gaussian quadrature numerical integrations may be preferred. Figure~\ref{fig:NumErrorTimeContourAnaQmax20} shows how the computation error and cost of numerical integration varies with the number of Gauss points and the order of spherical harmonics expansion used in the stress field expressions. The numerical integration error is evaluated with respect to an analytical traction force integration using $q_{max} = 20$ for Ni with a dislocation--interface distance of 100~$b$. As can be seen in Fig.~\ref{fig:NumErrorTimeContourAnaQmax20}, there are optimized combinations of integration point number and spherical harmonics expansion order to achieve a desired level of accuracy, and it is often more efficient to increase the order of spherical harmonics expansion than the number of Gauss integration points.

\section{Concluding remarks}
Using spherical harmonics expansions of the derivatives of Green's function, we constructed the expressions for the interface traction force exerted by dislocation stress field in anisotropic elastic media, and develop hierarchical recurrence relations to integrate the spherical harmonics series to calculate the traction force. It is found that all the triples integrals associated with the spherical harmonics are functions of a few analytically solvable seed integrals. Compared with numerical integrations of the traction force, our analytical integrations have gained substantially in computation precision and speed. This development of analytical interface traction integrations can impart accuracy and efficiency to DD--FE/BE simulations of the elastic interactions between dislocations and interfaces in general elastically anisotropic crystalline materials.

\section*{Acknowledgments}
We thank Sylvain Queyreau for helpful discussions. This work was performed under the auspices of the U.S. Department of Energy by Lawrence Livermore National Laboratory under Contract DE-AC52-07NA27344. Research was sponsored by the Army Research Laboratory and was accomplished under Cooperative Agreement Number W911NF-12-2-0022. The views and conclusions contained in this document are those of the authors and should not be interpreted as representing the official policies, either expressed or implied, of the Army Research Laboratory or the U.S. Government. The U.S. Government is authorized to reproduce and distribute reprints for Government purposes notwithstanding any copyright notation herein.

\bibliographystyle{paperRef}
\bibliography{paperRef}

\end{document}